\documentclass{sf2a-conf}
\usepackage{graphicx}
\begin{document}
\TitreGlobal{SF2A 2008}

\title{Formation and Properties of Molecular Cloud Cores}
\author{Dib, S.$^{1,}$}\address{Service d'Astrophysique, DSM/Irfu, CEA/Saclay, F-91191 Gif-sur-Yvette Cedex, France; email: sami.dib@cea.fr}
\address{Lebanese University, Faculty of Sciences, Department of Physics, El Hadath, Beirut, Lebanon}
\author{Galv\'{a}n-Madrid, R.$^{3,}$}\address{Harvard-Smithsonian Center for Astrophysics, 60 Garden Street, Cambridge, MA 02138, USA; rgalvan@cfa.harvard.edu}
\address{Centro de Radioastronom\'{i}a y Astrof\'{i}sica, Universidad Nacional Aut\'{o}noma de M\'{e}xico, 58089, Morelia, Michoac\'{a}n, Mexico; e.vazquez@astrosmo.unam.mx}
\author{Kim, J.}\address{Korea Astronomy and Space Science Institute, 61-1, Hwaam-dong, Yuseong-gu, Daejeon 305-348, Korea; jskim@kasi.re.kr}
\author{V\'{a}zquez-Semadeni, E.$^{4}$}
\runningtitle{Formation and Properties of Molecular Cloud Cores }
\setcounter{page}{237}
\index{Dib, S.}
\index{Galvan-Madrid, R.}
\index{Kim, J.}
\index{Vazquez-Semadeni, E.}

\maketitle
\begin{abstract} 
In this paper, we review some of the properties of dense molecular cloud cores. The results presented here rely on three-dimensional numerical simulations of isothermal, magnetized, turbulent, and self-gravitating molecular clouds (MCs) in which dense core form as a consequence of the gravo-turbulent fragmentation of the clouds. In particular we discuss issues related to the mass spectrum of the cores, their lifetimes and their virial balance. 
\end{abstract}

\section{The Simulations}
We performed 3D numerical simulations of magnetized, self-gravitating, and turbulent isothermal MCs using the TVD code (Kim et al. 1999) on grids with $256^{3}$ and $512^{3}$ cells (V\'{a}zquez-Semadeni et al. 2005; Dib et al. 2007a; Dib et al. 2008a). The basic features of these simulations are: Turbulence is driven until it is fully developed (at least for 2 crossing timescales) before gravity is turned on. The Poisson equation is solved to account for the self-gravity of the gas using a standard Fourier algorithm. Turbulence is constantly driven in the simulation box following the algorithm of Stone et al. (1998). The kinetic energy input rate is adjusted such as to maintain a constant rms sonic Mach number $M_{s}=10$. Kinetic energy is injected at large scales, in the wave number range $k=1-2$. Periodic boundary conditions are used in the three directions. In physical units, the simulations have a linear size of 4 pc, an average number density of $\bar{n}=500$ cm$^{-3}$, a temperature of 11.4 K, a sound speed of $0.2$ km s$^{-1}$, and an initial {\it rms} velocity of 2 km s$^{-1}$. The Jeans number of the box is $J_{box}=4$ (number of Jeans masses in the box is $M_{box}/M_{J,box}=J_{box}^{3}=64$, where $M_{box}=1887$ M$_{\odot}$). The simulations vary by the strength of the magnetic field in the box with $B_{0}= 0, 4.6, 14.5$, and 45.8 $\mu$G for the non-magnetized, the strongly supercritical, the mildly supercritical, and the subcritical cloud models, respectively. Correspondingly, the plasma beta and mass-to-magnetic flux (normalized for the critical value for collapse $M/\phi =(4 \pi^{2} G)^{-1/2}$; Nakano \& Nakamura 1978) values of the box are $\beta_{p,box}=\infty, 1, 0.1$, and $0.01$, and $\mu_{box}=\infty, 8.8, 2.8$, and $0.9$, respectively. Cores are identified using a clump-finding algorithm that is based on a density threshold criterion and a friend-of-friend approach as described in Dib et al. (2007a). We restrict our selection of cores to epochs where the Truelove criterion (Truelove et al. 1997) is not violated in any of them. Thus, the derived properties of our numerical cores can be best compared to those of starless prestellar cores.
   
\section{Core Mass Function}
\vspace{6.6cm}
\begin{figure} [h]
\centering
\begin{minipage} [h] {7.5cm}
\begin{picture} (6,6) \includegraphics[width=7.4cm]{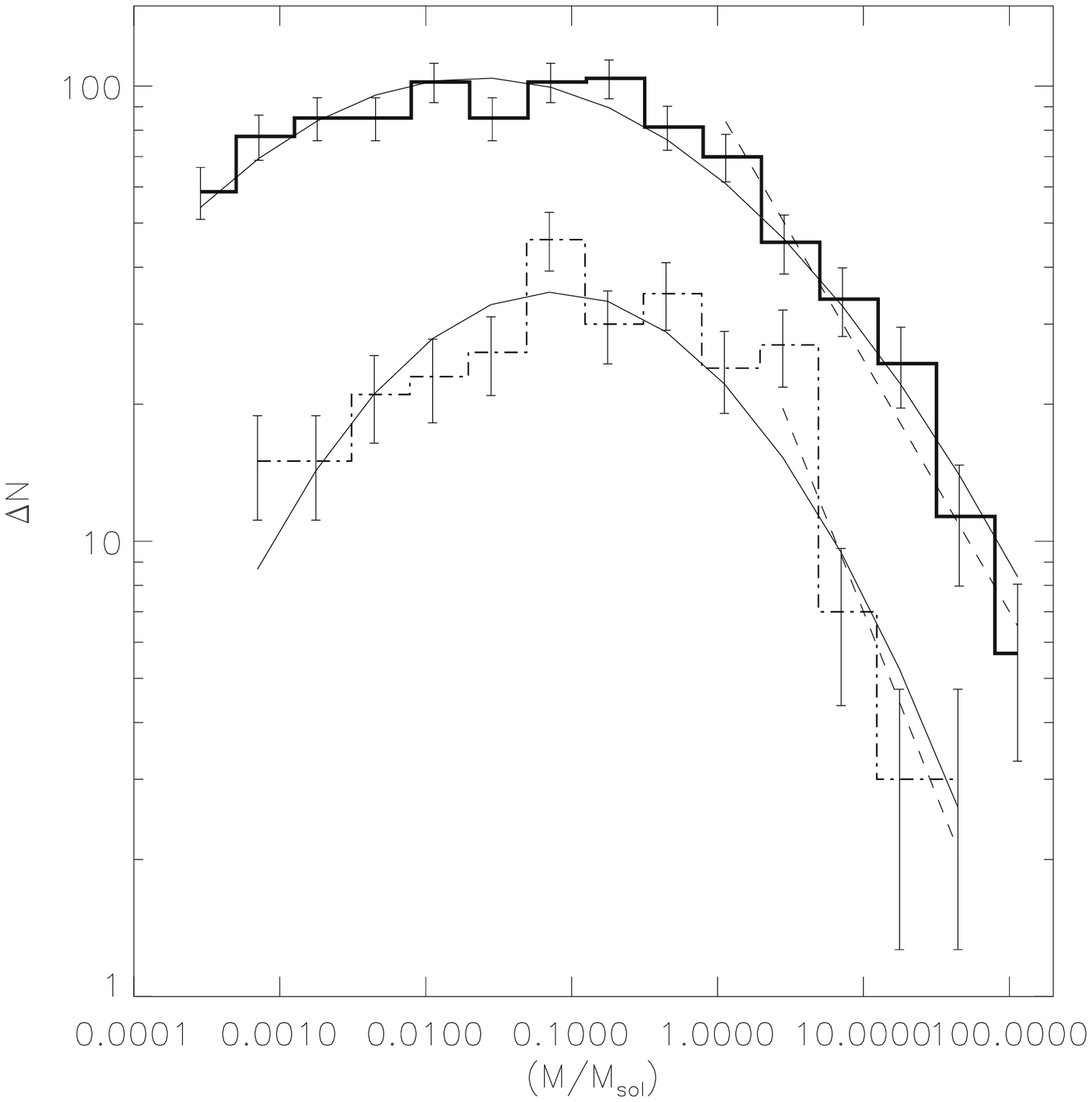} \end{picture}\par
\caption{\footnotesize Core mass function for cores identified at density thresholds of 20 and 50 times the average density (full thick and dot-dashed line respectively). Over-plotted to the CMF is a lognormal fit (full line) and power law fits in the intermediate to high mass regimes (dashed lines).} 
\label{fig1}
\end{minipage}
\begin{minipage}[h] {7.5cm}
\begin{picture} (6,6) \includegraphics[width=7.4cm]{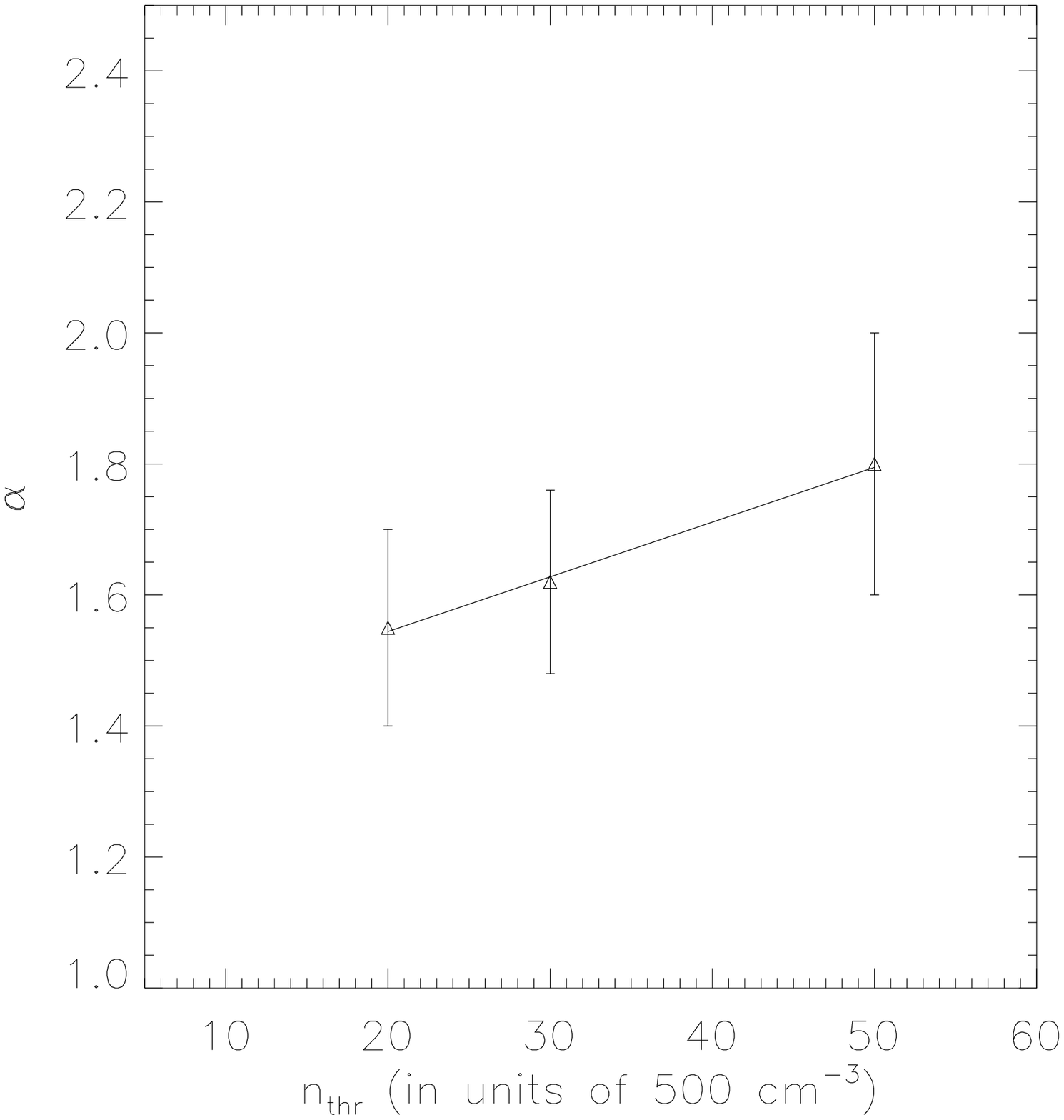} \end{picture}\par
\caption{\footnotesize Slope of the CMF in the intermediate and high mass regimes as a function of the density threshold. The slope of the CMF is steeper when cores are identified with higher density tracers/thresholds.}
\label{dib_fig2}
\end{minipage}
\end{figure}

One important area of modern astrophysics revolves around characterizing the initial mass function of stars (i.e. mass function of stars at their birth), and assessing whether it is universal (i.e. independent of environment) or not.  Early determinations of the IMF have been obtained by Salpeter (1955) who showed that the field star population can be described by a power law of the form $dN/dM=M^{\alpha}$ with $\alpha \sim -2.35$ in the intermediate to high mass regimes. Subsequent determinations of the IMF, particularly that of stellar clusters, have derived values of $\alpha$ that are in the range $\sim [-1.8,-2.7]$ (e.g., Massey et al. 1995; Sharma et al. 2007). A related important issue is that of the relationship between the IMF and the core mass function (CMF). Observations of dense cores in nearby star forming regions tend to indicate that their mass distribution is not very different from that of the IMF (e.g., Motte et al. 1998; Johnstone \& Bally 2006) which would indicate that the shape of the IMF might be already set in the early gaseous phase. Using our above described simulations, we have constructed the CMF at several density thresholds. Fig. 1 displays the CMF for cores identified at the density thresholds of $n_{thr}=20\bar{n}=10^{4}$ cm$^{-3}$ and $50\bar{n}=2.5\times 10^{4}$ cm$^{-3}$. The derived values of the slopes in the intermediate to high mass regimes show a steepening of the slope of the CMF with increasing threshold. Equivalently, the width of the CMF decreases with increasing threshold when the CMFs are fitted with a lognormal distribution.  Our statistics of cores does not allow us to probe the values of the slopes for cores identified at higher density thresholds, e.g., in the range of $200\bar{n}=10^{5}$ cm$^{-3}$ at which all cores are gravitationally bound. However, an extrapolation of the slope-threshold relation in Fig. 2 seems to indicate that values of the slopes in the range [-2.5,-3] can be expected for these threshold values. Nonetheless, it shoud be noted that the power law shape (or lognormal) of the CMF can be substantially modified as time evolves under the influence of various physical processes such as gas accretion and core coalescence (Dib et al. 2007b; Dib 2007; Dib et al. 2008a,b,c)   

\section{Virial Balance}
In Dib et al. (2007a) and Dib \& Kim (2007), we investigated the detailed virial balance of the cores formed in the simulations. By calculating all the terms of the virial theorem, we could show that the cores are dynamical, out of equilibrium structures. In each model, there is a mixed populations of gravitationally bound cores, cores that are bound by external compressions, and unbound dispersing objects.  We compare the diagnostic of the gravitational boundedness of cores using the detailed virial balance to that made using other simplified indicators, such as the virial parameter $\alpha_{vir}$ which compares the sum of the kinetic and thermal energy to gravity, the mass to magnetic flux ratio which compares the importance of magnetic support against gravity, and the Jeans number which compares the importance of thermal support against gravity. In all simulations, we find a trend in which, for gravitationally bound cores, the virial balance of such objects indicates that their inner parts (i.e., when cores are defined at high density thresholds) are more gravitationally bound than when the same objects are identified with lower density tracers (inner+outer parts). The simplified indicators seemed to indicate a different result in which the extended objects are more gravitationally bound than their inner parts. The discrepancy is nonetheless not very surprising considering the fact that the simplified indicators describe only a part of the energetic balance of the cores and that all of them neglect the surface energy terms that appear in the virial theorem. A comparison of the mass-virial parameter relation $\alpha_{vir}=M_{c}^{\beta}$ for the different simulated clouds (and at different epochs) with the observations (e.g. Bertoldi \& McKee 1992; Williams et al. 2000) shows that the closest agreement with the observations is obtained for the case of the mildly magnetically supercritical cloud model. The value of $\beta$ in this model is found to be $\sim -0.6$ (Fig. 3) in contrast to values of $\beta > -0.5$ for the strongly supercritical and non-magnetic cases. This seems to indicate that real molecular clouds in nearby star forming regions and in which cores are observed to form might be in a state of near magnetic criticality or slight super-criticality. 

\begin{figure}
\centering
\includegraphics[width=9.0cm] {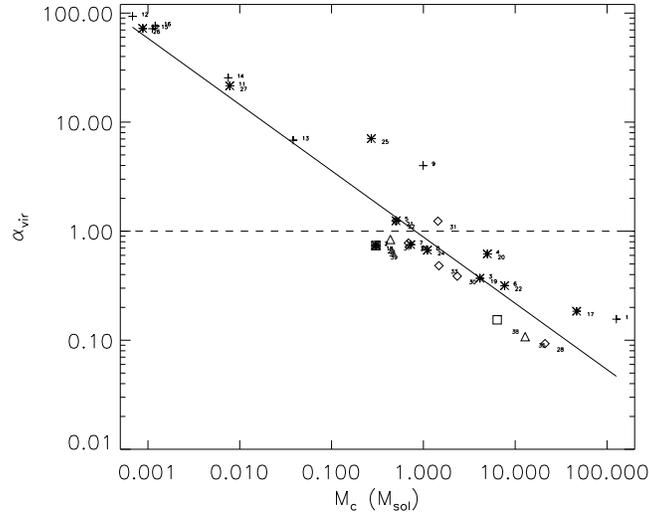}
\caption{The virial parameter as a function of the mass for the clumps and cores in the mildly magnetically supercritical simulation after $t=1.6$ Myr of turning on self-gravity. }
\label{dib_fig3}
\end{figure}

\section{Cores Lifetimes}
In V\'azquez-Semadeni et al. (2005) the time evolution of the cores was studied for simulations of different initial mass-to-magnetic flux ratio. Galv\'an-Madrid et al. (2007) performed mesurements of the core lifetimes as a function of the density threshold $n_{thr}$ used to define the cores for the likely case of MCs that are slightly magnetically supercritical (see above). The prestellar lifetimes of the cores ranged between a few to several free-fall times, with a mean value of  $\tau_{pre} \sim 6$ $t_{ff}$, where $t_{ff} \equiv (3\pi / 32G\rho)^{1/2}$.  Fig. 4 displays the lifetimes of the cores in the simulations against their threshold density (which is similar to the initial mean volume density). The observational data compiled by Ward-Thompson et al. (2007) is overplotted (filled squares). There is good agreement between the prestellar lifetimes of the simulations and the observations. Galv\'an-Madrid et al. (2007) also performed indirect estimations of the starless-to-protostellar core ratio. They found that it decreases from $\sim 5$ at $n_{thr} \sim 10^4$ cm$^{-3}$ to $\sim 1$ at  $n_{thr} \sim 10^5$ cm$^{-3}$, in rough agreement with observational estimates (e.g., Lee \& Myers 1999; Hatchell et al. 2007).
 
\begin{figure}
\centering
\includegraphics[width=10.2cm] {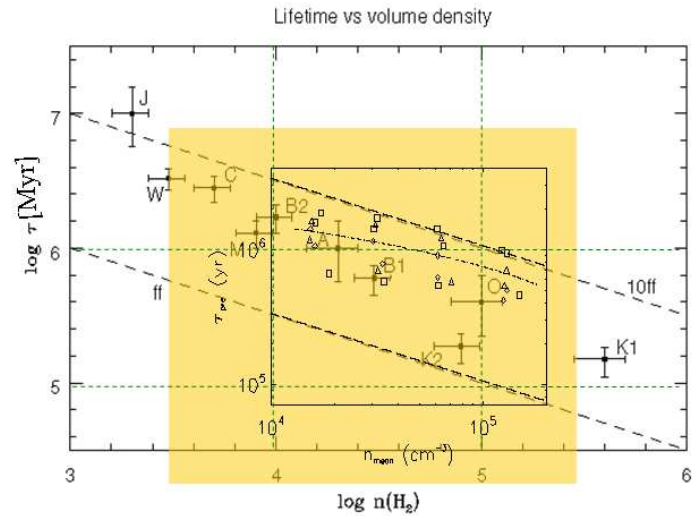}
\caption{Duration of the prestellar stage for the collapsing cores in the simulations as a function of their mean density. Diamonds, triangles, and open squares correspond to different runs. The loci of $\tau_{pre} = 1$ $t_{ff}$ and $\tau_{pre} = 10$ $t_{ff}$ are marked by the lower and upper long-dashed lines respectively. The linear fit to the simulation measurements ($\tau_{pre} \simeq 6$ $t_{ff}$) is marked by the dash-dotted line. Over-plotted to the simulations are observational data points (filled squares, taken from Ward-Thompson et al. 2007). }
\label{dib_fig4}
\end{figure}

\bigskip
S. D. would like to thank  the organizers of the SF2A-PCMI Dynamique workshop for the invitation to speak.

\end{document}